\documentclass{aa}  
\usepackage[utf8]{inputenc}
\usepackage{bm}
\usepackage{algpseudocode}
\usepackage{algorithm}
\usepackage{amsmath}
\usepackage{xcolor}
\usepackage{subfig}
\usepackage{comment}
\DeclareMathOperator*{\argmax}{arg\,max}

\newcommand{\action}{\boldsymbol{a}_t}

\newcommand{\st}{\boldsymbol{s}_t}

\usepackage{graphicx}
\usepackage{txfonts}
\usepackage{graphicx}
\usepackage{txfonts}
\begin{document}

   \title{On-sky demonstration of reinforcement learning for adaptive optics control}

   \subtitle{PO4AO on PAPYRUS at OHP}

    \author{J. Nousiainen
          \inst{1}
          \and V. Chambouleyron \inst{2}
          \and B. Neichel \inst{2}
          \and S. Cetre \inst{3}
          \and J-F. Sauvage \inst{2}
          \and A. Alagao \inst{2}
          \and M. Kasper \inst{1}
          \and J. Dray \inst{2,4}
          \and R. Fétick \inst{2}
          \and B. Engler \inst{1}
          }

   \institute{European Southern Observatory,  
   \email{jalo.nousiainen@eso.org} \and 
   Aix Marseille University, CNRS, CNES, LAM, Marseille, France \
   \and Wakea Consulting, 4 Rue Joseph Bouchayer, 38100 Grenoble \ 
   \and Bertin Alpao, 727 Rue Aristide Berges, 38330 Montbonnot-Saint-Martin, France
             }

   \date{Received March 9, 2026; accepted June 8, 2026}

  \abstract
     {Reinforcement learning (RL)-based algorithms have recently emerged as a promising approach for adaptive optics (AO) control. In simulations and laboratory experiments, they have demonstrated robustness to real-world effects such as photon and detector noise, misregistration, vibrations, and rapid variations in seeing conditions. However, their performance has not yet been validated on sky.}
     {We report the first on-sky demonstration of a reinforcement learning controller for adaptive optics, named Policy Optimization for AO (PO4AO). We further analyze its on-sky behavior and identify directions for improving the algorithm and its implementation.}
    {PO4AO was implemented and deployed on the Papyrus adaptive optics system installed at the Coudé focus of the 1.52 m telescope (T152) at the Observatoire de Haute-Provence (OHP). A Python-based implementation was interfaced with the existing real-time controller (DAO RTC) via shared-memory buffers. The performance of PO4AO was compared to that of a standard integrator controller over several nights, covering a range of flux levels and atmospheric conditions.}
   {PO4AO consistently outperformed the standard integrator in all tested configurations. The controller successfully learned and compensated for vibration patterns and demonstrated strong robustness to measurement noise. Once tuned for Papyrus, PO4AO operated in a turnkey fashion, using a single set of hyperparameters across varying observing conditions and science targets. These performance gains were achieved despite a non-optimized Python implementation introducing approximately 720~\textmu s of additional latency, along with control jitter and occasional frame drops.}
    {When properly implemented and optimized, PO4AO constitutes a robust and high-performance turnkey controller for single-conjugate adaptive optics systems, paving the way for broader adoption of reinforcement learning strategies in on-sky AO operations.}


   \maketitle
%
\section{Introduction}
Adaptive optics (AO) was first proposed nearly seventy years ago by \citet{babcock1953possibility}. Its first successful astronomical on-sky implementation was later demonstrated by \citet{merkle1989successful}, who restored diffraction-limited performance on the same 1.52-m telescope on which the results presented in this paper were obtained, 36 years later. Since that pioneering demonstration, AO has undergone major developments and now encompasses a broad range of astronomical applications. Modern systems are optimized for different scientific objectives, including wide-field correction, faint-object detection, multi-directional compensation, and high-fidelity correction over narrow fields of view.

In this paper, we focus on narrow-field correction systems, specifically classical single-conjugate adaptive optics (SCAO). These systems rely on a bright reference star located close to the science target to estimate atmospheric wavefront distortions along a nearly identical optical path. SCAO operates in closed loop: the wavefront sensor (WFS) measures the residual aberrations after correction by the deformable mirror (DM), and the control algorithm updates the DM commands to minimize the residual wavefront error.

SCAO systems are most commonly controlled using a linear reconstruction combined with an integrator control law \citep{roddier1999adaptive}, hereafter referred to as the integrator. This control architecture is simple, computationally efficient, and robust under a wide range of observing conditions. Nevertheless, it remains fundamentally limited by several non-negligible error sources, including temporal delay errors \citep{cantalloube2018origin}, dynamic misregistration between the DM and WFS \citep{heritier2018new}, and nonlinearities that are not captured by the calibrated interaction matrix \citep{verinaud2004nature}. These limitations have motivated extensive research into advanced AO control strategies over the past decade, including predictive control, model-based optimal control, and more recently, data-driven approaches.

Some research focuses on mitigating temporal error, for example, utilizing quadratic loss-based linear controllers (e.g., Kalman filters, \citealp{kulcsar2006optimal, paschall1993linear, gray2012ensemble, conan1a2011integral, correia2010adapting, correia2010optimal, correia2017modeling}), sometimes augmented with machine learning for system identification \citep{sinquin2020sky, haffert2021data, dinis2024upgrading}. Other techniques span from linear filtering methods to modal filtering in Fourier or Zernike bases \citep{guyon2017adaptive, poyneer2007fourier, dessenne1998optimization, van2017performance, van2019impact, nousiainen2024power}, as well as neural network–based prediction \citep{swanson2018wavefront, sun2017bayesian, liu2019using, wong2021predictive}. Closed-loop implementations have also been studied; for example, \citet{males2018ground} analyzed their impact on post-coronagraphic contrast using a semi-analytic framework, while \citet{swanson2021closed} employed supervised learning to train convolutional neural networks (CNNs) that compensate for temporal errors. Some of these concepts have already been validated on-sky (e.g., \citealp{van2022predictive}). Additionally, one class of algorithms aims to mitigate nonlinear errors. The methods vary from first-order corrections with optical gains (e.g., \citealp{deo2019telescope, chambouleyron2020pyramid}), to full model-based non-linear reconstruction (e.g., \citealp{hutterer2018nonlinear, hutterer2019real, chambouleyron2024using, frazin2018efficient}), and finally to machine learning utilizing techniques (e.g., \citealp{Landman:20, landman2024making, swanson2018wavefront, wong2023nonlinear, archinuk2023mitigating, weinberger2024transformer, perez2025open}). Moreover, \cite{landman2025making} tested CNN-based reconstruction on-sky with MagAO-X, demonstrating the feasibility (e.g., fast enough inference time) of Deep NN and improvement in performance for fainter stars. 

Moreover, in recent years, several reinforcement learning (RL)– based approaches (bypassing the need for labeled datasets) for astronomical adaptive optics systems with WFSs have been proposed \citep{Pou:22, pou2024integrating, landman2020self, landman2021self, nousiainen2021adaptive, nousiainen2022toward}. A comprehensive overview of machine learning methods for astronomical wavefront control and phase prediction is given by \cite{fowler2023tempestas}. In recent years, RL has also been explored in broader AO contexts. Representative applications include deep RL for aberration correction in wavefront-sensorless imaging, formulated as a Markov decision process \citep{ke2019self}; adaptive microscopy with deformable mirrors controlled by deep RL \citep{durech2021wavefront}; RL-driven AO for fiber-coupled optical communication links \citep{Parvizi2023}; and RL to correct piston misalignment between segments in a segmented mirror telescope \citep{guerra2020towards}. Additionally, RL-trained controllers have attracted increased attention for other complex real-world control problems, including nuclear fusion reactors, radio and space telescopes, particle accelerators, and gravitational-wave observatories \citep{buchli2025improving, degrave2022magnetic, yatawatta2021deep, xiong2020intelligent, kaiser2024reinforcement}. 

In this paper, we focus on a specific RL method, Policy Optimization for AO (PO4AO), which has shown promising, robust results across various numerical simulations and lab tests \citep{nousiainen2022toward, nousiainen2024laboratory, nousiainen2022advances, dray2024comparison, camelo2024reinforcement, camelo2023papyrus}.  PO4AO addresses temporal control and reconstruction in the XAO control loop as a single reinforcement learning problem. Consequently, PO4AO addresses various aspects of AO control, including predictive control, optimal gain compensation, misregistration identification, reconstruction algorithms, and vibration cancellation. This paper gives an on-sky demonstration of PO4AO. We implemented and adapted PO4AO to the PAPYRUS adaptive optics system \citep{muslimov2021current, striffling2023papyrus, fetick2023papyrus} on a 1.52m telescope, and tested it on various stars of different magnitudes on several nights. Compared to integrator control, PO4AO provides significant performance gains in all test cases. We demonstrate that PO4AO is capable of learning vibration patterns, predictive control, and is robust against WFS measurement noise. Moreover, once the hyperparameters are tuned for Papyrus, the use of PO4AO is entirely a turnkey operation — the controller is operated with the same set of hyperparameters for each condition and science target, and no additional tuning is required.

The paper’s structure is as follows: In Section \ref{sec:papyrus}, we introduce the Papyrus AO system and its reference controller. Section \ref{sec:po4ao} gives a short description of PO4AO; details can be found in the previous publications \citep{nousiainen2024laboratory, nousiainen2022toward}. Section 3 also describes how the method is implemented on Papyrus. In Section \ref{sec:results}, we present the results of experiments during three different nights. The results include recorded point-spread functions (PSFs) and analyzed telemetry data. Section \ref{sec:analysis} presents a brief linear analysis of the behavior of PO4AO, and Section \ref{sec:conclusion} discusses the findings and future work.

\section{The Provence Adaptive Optics PYramid Run System--PAPYRUS}\label{sec:papyrus}

\begin{figure}[ht]
\centering
\includegraphics[trim={2.cm 12cm 2cm 2cm}, clip, width=0.48\textwidth]{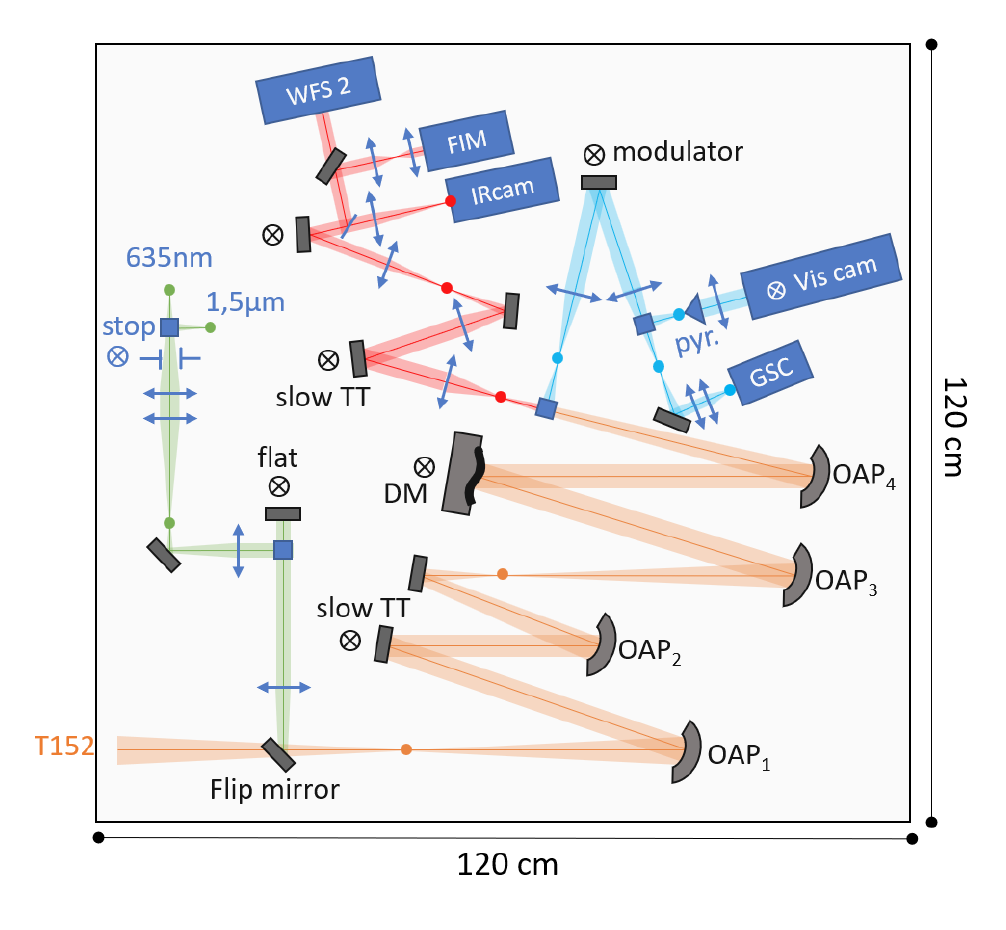}
    \caption{Schematic diagram of the PAPYRUS bench during the period of gathering for this paper. The bench consists of four key-subystems: the dual calibration unit (green), the common path (orange), the visible wavefront sensing path (blue), and the near-IR path (red). The focal planes are marked as solid circles while the pupil planes are represented by a circle with a cross inside.  The acronyms are as follows: off-axis parabola (OAPs), tip-tilt mirrors (TTs), deformable mirror (DM), gain-scheduling camera (GSC) and fiber injection module (FIM)}
    \label{fig:papyrus}
\end{figure}

The Provence Adaptive Optics PYramid Run System (PAPYRUS) is a pyramid-based AO instrument installed at the Coudé focus of the 1.52 m telescope (T152) at OHP. The AO instrument serves as a development platform for new AO concepts and ideas, and as a pathfinder for the challenging AO systems of the upcoming Extremely Large Telescopes (ELTs).

PAPYRUS consists of four sub-systems: the dual-wavelength calibration path, the common path, the visible wavefront sensing path, and the infrared path. Light from the Coudé focus of the T152 is directed to the PAPYRUS bench through a periscope consisting of folding mirrors. It is focused at a location within the bench that coincides with the light source's focus in the calibration branch. Switching between the calibration source and on-sky is done through a flip mirror located before the telescope focal plane. At a distance of one focal length, an off-axis parabola (OAP1) collimates the beam and directs it to a slow tip-tilt mirror in the pupil plane. From the mirror, the beam is reimaged by two OAPs (OAP2 and OAP3) and a flat mirror onto the DM, which serves as the system's aperture stop. The deformable mirror (DM), manufactured by ALPAO, consists of 241 actuators arranged on a 17 × 17 Cartesian grid. The optical design projects the telescope pupil onto the DM, resulting in a slightly smaller pupil than the reflective surface, yielding an effective 15 × 15 DM on the sky.  Light reflected by the DM is focused on OAP4. A dichroic filter located right before the focus reflects wavelengths shorter than 950 nm to the wavefront sensing branch and transmits the longer wavelengths to the infrared branch.  The visible wavefront sensing branch consists of a four-sided pyramid equipped with a modulation mirror and an OCAM2K from First Light Imaging company as the detector.  The detector is a $240 \times 240$ grid, and each pyramid pupil is approximately 70 pixels across. This results in an oversampling of roughly 4 pixels per DM actuator.

The infrared (IR) branch consists of a second slow TT in the pupil plane, followed by a series of lenses and mirrors that reimage the pupil and redirect the beam onto the surface of another flat mirror, which serves as a temporary placeholder for a second DM.  Light reflected by this mirror encounters a second dichroic filter, where wavelengths shorter than 1500 nm are transmitted and focused onto an IR camera (a CRED2), while the longer wavelengths are transmitted and reflected by another flat mirror towards the fiber injection module. This flat mirror stands on a magnetic base and can be removed to give way for tests on a future second wavefront sensor. Our experiment use only the CRED2 camera and no NIR WFS was deployed.

PAPYRUS can be operated during the daytime using its dual-calibration unit, composed of fiber-fed laser diodes at 635 nm and 1550 nm. These sources are reimaged at the telescope's focal plane by a series of lenses and mirrors.

As an adaptive optics platform, PAPYRUS served as a testbed for various wavefront sensing concepts (modulated and unmodulated four-sided pyramid, three-sided pyramid, gain scheduling camera), wavefront controllers and reconstructors, and various RTC hardware and software (Matlab, ALPAO RTC, DAO RTC, CPU \& GPU).  Utilizing the Durham Adaptive Optics real-time controller (DAO-RTC, \citealp{barr_2025_17264152}) as its baseline, PAPYRUS has been used to investigate calibration and optical gain-sensing techniques \citep{2025A&A...703A.253S} developed for HARMONI. The shared memory architecture of the DAO-RTC has enabled efficient integration and testing of NN-based approaches to address the nonlinear behavior of pyramid wavefront sensors \citep{2024SPIE13097E..0SW}. Furthermore, PAPYRUS delivers AO-corrected wavefronts to single-mode fiber-fed spectrographs, with dedicated modules available in the visible for RISTRETTO \citep{lovis2024ristretto} and in the infrared for compact fiber-fed spectrographs such as VIPA \citep{carlotti2022sky}, and SWIFTS \citep{bonneville2013swifts}.

The default controller of Papyrus is a leaky integrator. The interaction matrix $D$ is obtained by poking the actuators and recording the corresponding intensity changes in the WFS Camera. The interaction matrix is then inverted to a reconstruction matrix $C$ by decomposing into a smaller-dimensional Karhunen-Loève modal basis \citep{1994ESOC...48..187G}.

At a given time step $t$, the WFS measures the residual wavefront. The leaky integrator then obtains the new control voltages $\bm v_t$ from
\begin{equation}
\label{eq:lin_integrator}
\bm v_t = \alpha \bm v_{t-1} + g C_m \Delta \bm w_t,
\end{equation}

As discussed in the introduction, we compare PO4AO against the integrator. In Section \ref{sec:results}, we present results for two integrators: one implemented with the same Python interface as PO4AO and the other implemented with DAO RTC. Both integrators operate with the same reconstruction matrix, fixed single-gain value, and leakage. The DAO RTC integrator has lower latency and jitter. However, the Python integrator enables faster switching between Python-based PO4AO and the integrator.

\section{Policy optimization for Adaptive Optics}\label{sec:po4ao}
Policy optimization for Adaptive Optics (PO4AO) contains two learned parametric models: the dynamics model and the policy model. The dynamics model $\hat p_\omega(\cdot,\cdot)$, represented by a fully Convolutional NN (CNN), predicts the subsequent WFS data reconstruction ($\bm o_{t+1} = C_m \Delta \bm w_t,$) given the action (residual command) to be applied $\bm a_{t}$, and telemetry history, the "state", $\bm s_t$, containing previous reconstructions ("observations"; $\bm o_t, \bm o_{t-1}, \cdots$), previous "actions', i.e., residual DM commands noted as ($\bm a_t, \bm a_{t-1}, \cdots$). That is,
\begin{equation*}
\tilde {\bm o}_{t+1} =  p_\omega(\bm o_t, \bm o_{t-1}, \cdots, \bm o_{t-k}, \bm a_t, \bm a_{t-1}, \cdots, \bm a_{t-k}).   
\end{equation*}

The Policy ($\pi_\theta(\cdot)$, also presented as a CNN) is a nonlinear control law that maps the state to a new DM command vector. 
\begin{equation}
\bm a_t = \pi_\theta(\bm s_t) = \pi_\theta(\bm o_t, \bm o_{t-1}, \cdots, \bm o_{t-k}, \bm a_{t-1}, \cdots, \bm a_{t-k}).   
\end{equation}
The action $\bm a_t$ is then added on top of the last DM command with a leakage, that is, the full DM command is given by $\bm v_t = \alpha \bm v_{t-1} + \bm a_t$. 

The dynamics model is trained using a continuously collected closed-loop telemetry dataset $\mathcal{D}$. The policy is then trained with the dynamics model utilizing the same data. More precisely, we sample starting states from the dataset, let the policy decide actions, and simulate trajectories (of length defined by planning horizon $H$) using the dynamics model. We collect the reward (performance criteria) at each time step and backpropagate the errors to obtain gradients for the policy model parameters. Let us define the reward as the negative squared Euclidean norm  of the reconstructed wavefront:
\begin{equation}
    \hat r_\omega(\st,\action) = - \|\tilde {\bm o}_{t+1}\|^2,
\end{equation}
where $\tilde {\bm o}_{t+1}$ is obtained from $\tilde {\bm s}_{t+1} = \hat p_\omega(\st, \action)$.

The policy optimization is then given by
\begin{equation}
    \label{eq:optimization_of_hatr}
    \argmax_\theta \sum_{{\bf s} \in \mathcal{D}}  \sum_{t=1}^{H} \hat r_\omega(\tilde \st, \pi_\theta(\tilde \st)),
\end{equation}
where $H$ is the planning horizon and
\begin{equation*}
    \tilde {\bf s}_1 = {\bf s} \quad \text{and} \quad \tilde {\bf s}_{t+1} = \hat p_\omega(\tilde \st, \pi_\theta(\tilde \st)).
\end{equation*}
Both models are trained in parallel with the control loop, enabling them to continuously adapt to prevailing atmospheric conditions. The complete algorithm, practical details, and the NN designs (including the number of layers, filters and activation functions), are described by \cite{nousiainen2024laboratory}.

\subsection{PO4AO parameters}
PO4AO has many hyperparameters that can be tuned; however, the default values used in \citep{nousiainen2024laboratory} worked well in most cases, and we needed to tune only a few. The most important parameter to tune is "warm-up noise" level (see \citealp{nousiainen2024laboratory}). Which presents the maximum level of noise injection during warm-up data collection. During the warm-up (first 20 episodes), PO4AO runs the integrator control with an added noise component (after each episode, the noise level is linearly reduced) to produce an initial dataset on which the first NN (policy and dynamics) models are trained. After the warm-up, the PO4AO runs without any additional noise component (i.e., exploration noise). It is crucial that warm-up dataset contains both "good" and "bad" controls, that is, observation-action pairs that yield large and small WFS measurements. The integrator controller without noise injection would only provide "good" control, close to optimal. The parameter should be chosen so that only the first few episodes are dominated by noise injection ("bad control"), and the last episodes are run with effectively no noise (pure integrator, "good" control). After this parameter was tuned, we could close the loop with PO4AO on default values.

We then further tuned several additional parameters. We set the episode length to 1000 steps (corresponding 2~sec for the framerate of 500Hz) to reduce model update frequency and, consequently, frame skipping. We set the history length to 64, which delivered the best performance trading inference time against the ability to identify vibrations. The number of controlled modes was set to the same value as in the Papyrus reference integrator, namely 209. The key PO4AO parameters are noted in Table \ref{table:parameters}.

\subsection{DAO interface, PO4AO implementation and computational cost}
The Python-based PO4AO implementation is composed of two parallel threads: the control thread and the training thread. We interface the DAO RTC pipeline with the PO4AO control thread via shared memory buffers. In this setup, the DAO RTC pipeline acquires raw WFS images, processes them, applies the flat reference, and multiplies the result by the reconstruction matrix. The resulting "delta voltages" (i.e., the linear wavefront reconstruction, $\bm o_t$) are then written to the shared memory buffer, after which the DAO RTC loop pauses. A control thread subsequently reads these delta voltages ($\bm o_t$) from the shared memory. The control thread can also maintain a history of previous residual voltages and applied actions (i.e., $(\bm o_{t-1}, \cdots, \bm o_{t-k}, \bm a_{t-1}, \cdots, \bm a_{t-k})$). Both the stored history (past delta voltages and actions) and the newly received delta voltages are passed to a Policy NN, which produces a set of new DM commands. These commands are written to the DM command buffer, where the DAO RTC retrieves them and continues for execution. Meanwhile, the control thread records the observed delta voltages and the corresponding DM commands into replay buffers (i.e., data sets, see \cite{nousiainen2024laboratory}). The training thread then accesses the replay buffers (data) and alternates between the dynamics model training loop and the Policy model training loop. Each time the Policy training circle completes, the training thread notifies the control thread of the Policy update, and the control thread uploads the new Policy parameters.  

The full DAO integrator latency (estimated from the empirical transfer function) is around 1.7 steps at 500Hz. The PO4AO and Python-integrator both add latency on top of this inherited delay. The DAO RTC is equipped with a single GPU; therefore, both the training and control threads share the same GPU for NN operations. This leads to significant jitter in the measured control latency of PO4AO, which was recorded at 720 µs with a standard deviation of 480 µs. At 500Hz loop speed, approximately 5\% of frames were skipped. The loop jitter in the latency was considerably lower when the training thread was stopped, suggesting that installing an additional GPU card (for the training thread) could stabilize the latency. However, the best results would require using a lower-level programming language, which would reduce latency to the raw policy NN forward pass (<300 \textmu s). The additional latency of Python-integrator was around 500 µs lower, and also the jitter was much smaller because the training thread was not running. A comprehensive overview of the latency terms for Python-based PO4AO using shared-memory buffers is provided in \cite{nousiainen2024laboratory}.

During each episode, the control thread updates the NN models. Each update circle, with the given parameters (e.g., number of gradient steps, history size), took 0.73 sec for $17 \times 17$ Papyrus DM. This execution time fit well with the episode length of 2 sec (1000 frames at 500Hz) and thus required no implementation optimization. However, we did not further experiment with the optimal update frequency, which might be faster, especially on gusty or versatile wind conditions. In this case, PO4AO might benefit from optimized training thread implementation, for example, parallelized for-loops over the dynamics model ensemble or Just-in-time compilation of PyTorch code.

Looking ahead to ELT-era instruments equipped with 5,000–10,000 actuator deformable mirrors (DMs), the computational cost of neural network (NN) inference and training increases significantly. While the convolutional layers in both the dynamics and policy networks are relatively shallow, highly parallelizable, and contribute only marginally to the overall runtime, the policy network contains a modal filter layer that performs a matrix multiplication from actuator space to actuator space. For high-order DMs, this operation begins to dominate the computational cost, particularly within the training thread. For example, the training cycle on the Papyrus $17 \times 17$ DM takes approximately 0.76 seconds. Using the same hyperparameters and NN architecture, the training cycle on a $128 \times 128$ DM—representative of ELT-scale extreme adaptive optics instrument takes about 4.4 seconds on a comparable GPU. This implies that achieving a similar adaptation speed at ELT scale will require further optimization and more powerful hardware.

\begin{table}[ht]
    \caption{Papyrus, Seeing, RL parameters}
    \label{table:parameters}
\begin{tabular}{ l l l} 
 \hline\hline
 \multicolumn{3}{c}{Telescope} \\
 \hline
         Parameter  & Value  &  Units  \\
 \hline
 Telescope diameter  &  1.52  & m \\
 Loop frequency   &  500  & Hz   \\
 Number of actuators   &  241   & -  \\
 Science Wavelength  &  1450 & nm \\
 WFS Wavelength  & 400-900 & nm \\
 Delay & ~3  & frames @ 500Hz\\
 P-WFS modulation & 5 & $\lambda$  D \\
 \hline
  \multicolumn{3}{c}{Observed targets V-band magnitudes} \\
 \hline
 Vega              &  0.09  &  - \\
 HD 177809         &  5.72  &  - \\
 Cygni (HR 8004)   &  6.66  &  - \\
 $\alpha$ Cas      &  2.20  &  - \\
  \hline 
\multicolumn{3}{c}{Seeing estimate during observing} \\
 \hline
 1. night, Vega             &  3.0  & arcsec \\
 2. night, Vega             &  2.8  & arcsec \\
 2. night, HD 177809        &  2.2  &  arcsec \\
 2. night, Cygni (HR 8004)  &  2.6  &  arcsec \\
 3. night,$\alpha$ Cas      &  3.8  &  arcsec \\
  \hline 
  
 \multicolumn{3}{c}{Modified PO4AO parameters} \\
 \hline
 Planning horizon ($H$)   &  4   & steps     \\ 
 Controlled KL modes      &  209 & -         \\
 Episode length      &  1000 & frames         \\
 Warm up episodes & 20 & episodes              \\
 Max warm-up noise & 5 & \% of full stroke \\
  \hline 
\end{tabular}
\end{table}

\section{Results}\label{sec:results}
This section reports the results we obtained on three nights: 6/18/2025 (1st night), 6/20/2025 (2nd night), and 9/11/2025 (3rd night). We recorded multiple datasets, and here we present the most informative data to illustrate various properties of the algorithm. On the first night, we observed strong chaotic vibrations due to a malfunctioning telescope tracking motor; therefore, we selected data from bright targets and focused on demonstrating PO4AO's vibration-control capabilities; see Sec. \ref{sec:1st}. 
On the second night, the vibration problem was resolved by maintaining the tracking motor hardware, and we selected data to demonstrate PO4AO's noise-robustness and turnkey nature by observing targets of varying magnitudes; see \ref{sec:2nd}. On the 1st and 2nd nights, we compare PO4AO results with those of the Python integrator, with an integrator gain of 0.4 (likely slightly smaller than the optimal value for bright targets). On the third night, we compared PO4AO with the internal, low-latency integrator and manually tuned the gain before recording data. We also fixed bugs in the telemetry recording function and obtained higher-quality telemetry data with fewer skipped frames; hence, we performed a simple linear analysis of PO4AO behavior on this data (see Sec. \ref{sec:analysis}). 

At the beginning of each night, we calibrated PO4AO by running the warm-up procedure as described by \cite{nousiainen2024laboratory} and recording the related warm-up data buffer. After this initial training procedure, the loop is closed using the most recent PO4AO policy model. The policy is updated every 2 seconds during execution, enabling the online adaptation to prevailing conditions. Each collected dataset contains 30.000 telemetry frames (WFS measurements and dm commands) and a set of science camera frames recorded during telemetry collection (500-5000 frames, i.e, $\sim$ 1 min, depending on the frame rate). The following metrics are used to evaluate the correction performance of PO4AO and the integrator controller:

\begin{enumerate}
    \item Strehl ratio: We compute the Strehl with MAOPY PSF model \citep{2019A&A...628A..99F}; see Figs. \ref{fig:first_night_psfs}, \ref{fig:second_night_psf} and \ref{fig:third_night_psfs}. We note that estimating on-sky Strehl is challenging, and the Strehl values obtained are only comparable across methods under similar seeing conditions. 
    \item Relative peak intensity To add redundancy to Strehl estimation, we take the integrator peak intensity and divide it by the PO4AO peak intensity for each data set; see Table \ref{table:results}.
    \item Variance of KL modes: We compute the variance over each mode of the WF residual measurements in the telemetry data.  See Figs. \ref{fig:1st_telemetry}, \ref{fig:2nd_telemetry} and \ref{fig:3rd_telemetry}. We chose variance over mean-squared error because DM saturation (especially in bad-seeing cases) sometimes introduces a bias term for lower-order modes in both the integrator and PO4AO.
    \item Temporal power spectral density of "tilt" mode: the PSD is extracted from the data using Welch's method.
\end{enumerate}

\subsection{First night, 6/18/2025: vibration control}\label{sec:1st}

\begin{figure}[ht]
\centering
\includegraphics[width=0.48\textwidth]{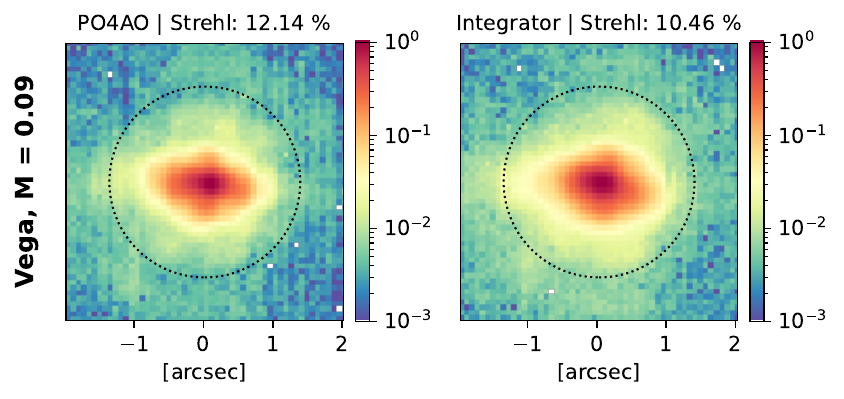}
    \caption{PSF during the 1st night, under the strong vibration. Left image: PO4AO (12.14\% Strehl), Right image: Python Integrator (10.46\% Strehl)}
    \label{fig:first_night_psfs}
\end{figure}

\begin{figure}
\centering
\subfloat[\label{fig:1st_var}]
        {\includegraphics[trim={0.cm 0cm 0cm 0cm}, clip, width=0.45\textwidth]{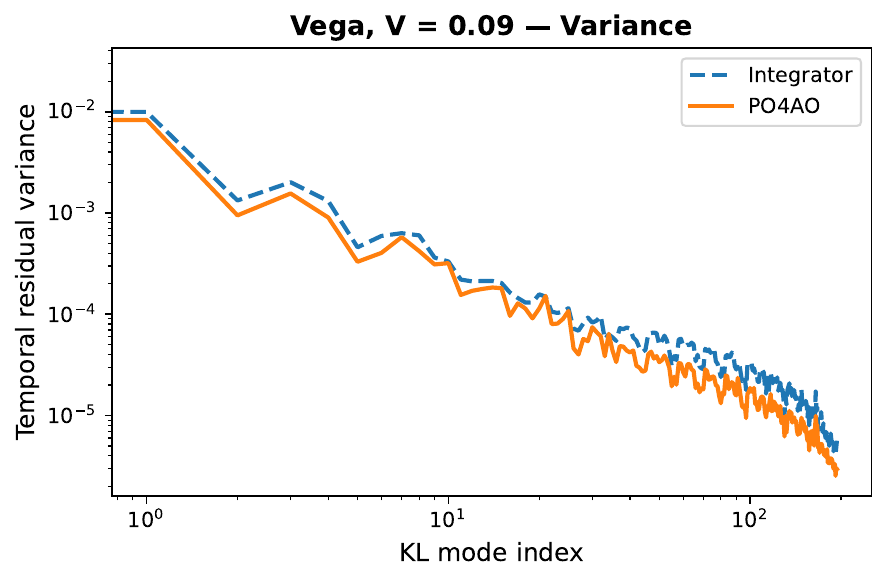}}

\subfloat[\label{fig:1st_psd}]
         {\includegraphics[trim={0.cm 0cm 0cm 0cm}, clip, width=0.45\textwidth]{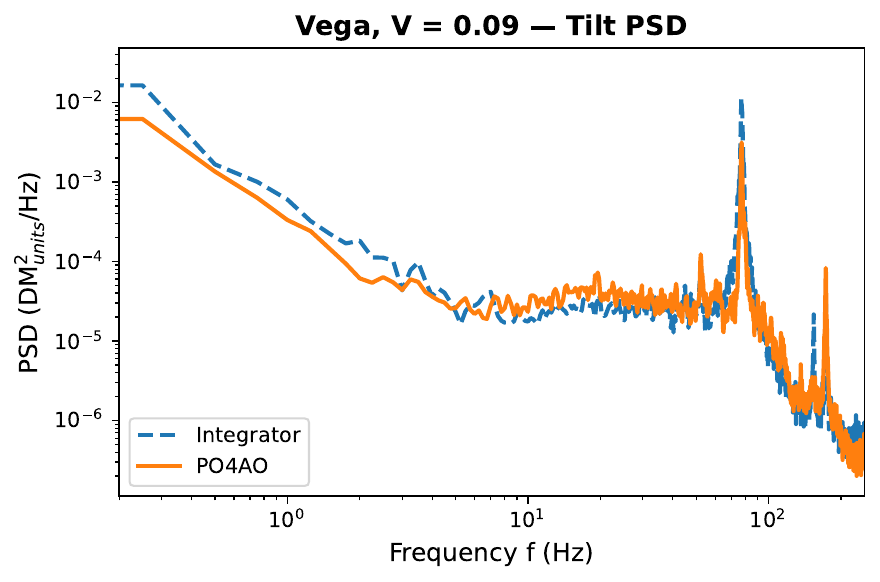}}
     
\caption[First night]
        {First night telemetry analysis. Top: Residual variance per KL mode. Bottom: temporal PSD of the residual phase.}
    \label{fig:1st_telemetry}
\end{figure}

On the first night, we closed the loop with both the integrator and PO4AO multiple times on multiple different targets. Here we present data on the brightest target, the 0th-magnitude "Vega". We chose to present the best Strehl-ratio case from both PO4AO and the integrator, across various options within a short time period (~15min). PO4AO yielded higher Strehl than the best Integrator across all datasets. We plot the long-exposure PSFs (Fig. \ref{fig:first_night_psfs}), the temporal modal variance of the WFS measurement modes, and the temporal PSD of the "tilt" mode (where the vibration was most visible) in Fig. \ref{fig:1st_telemetry}. The seeing estimates are given in Table \ref{table:parameters}.

As noted previously, on the 1st night, we observed a strong chaotic vibration shown by the spike in the temporal PSD in Fig \ref{fig:1st_psd}. This vibration leads to the horizontally elongated PSFs in Fig. \ref{fig:first_night_psfs}. The vibration was irregular, making it difficult to control (see Appendix A).

Fig. \ref{fig:first_night_psfs} compares the best PSFs. The non-common-path (NCPA) errors, including strong astigmatism, were not corrected. The PO4AO PSF is sharper, particularly in the vertical (up-down) direction, where the vibration was strongest, and the Strehl estimation was slightly better for PO4AO (12\%) than for the integrator (10\%).

Fig. \ref{fig:1st_telemetry} plots the temporal variance of the WFS measurements for each controlled mode and the temporal PSD of the "tilt" mode. From Fig. \ref{fig:1st_var}, we observe that PO4AO has lower residual variance across almost all KL modes. The largest gain in residual variance (a factor of 2.3) occurs at mode 1 ("tilt"), where we also observe the strongest vibration spike (see Appendix for relative gain plot). Otherwise, the biggest gains are in the mid spatial frequencies (from 50-150), where the relative gain is around 1.8. For other lower-order modes, the gain factor is around 1.2-1.6. Fig. \ref{fig:1st_psd} shows the temporal PSD of the tilt model. PO4AO reduces the vibration peak at 80 Hz by a factor of 4.5. There is also a spike at 150~Hz, where PO4AO reduction is a factor of 8, and around 175Hz, where the amplitude is the same for the integrator and PO4AO, suggesting no correction for this vibration. The reduction in the vibration peak was similar across the datasets, and this dataset was chosen to represent the average case.

\subsection{Second night,  6/19/2025: performance on different guide star magnitudes}\label{sec:2nd}

\begin{figure}[ht]
\centering
\includegraphics[width=0.48\textwidth]{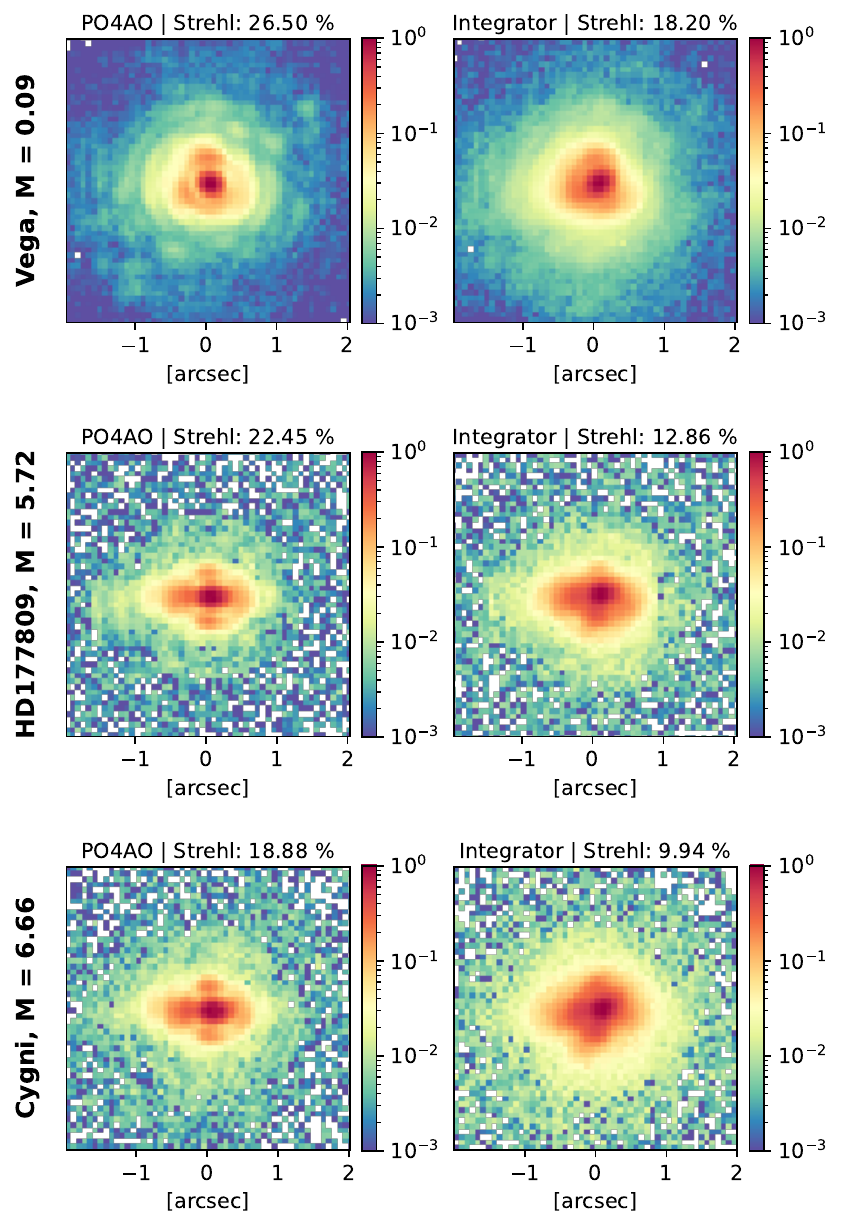}
    \caption{PSF during the second night. Each row compares the best PO4AO PSF and the best integrator PSF for the same target. The top row is Vega, the middle row is HD177809, and the bottom row is Cygni. The NCPA errors are corrected only for the Vega. We note that the integrator gain was probably not optimal at the given conditions and brightness.} 
    \label{fig:second_night_psf}
\end{figure}

\begin{figure}[ht]
\centering
\includegraphics[width=0.48\textwidth]{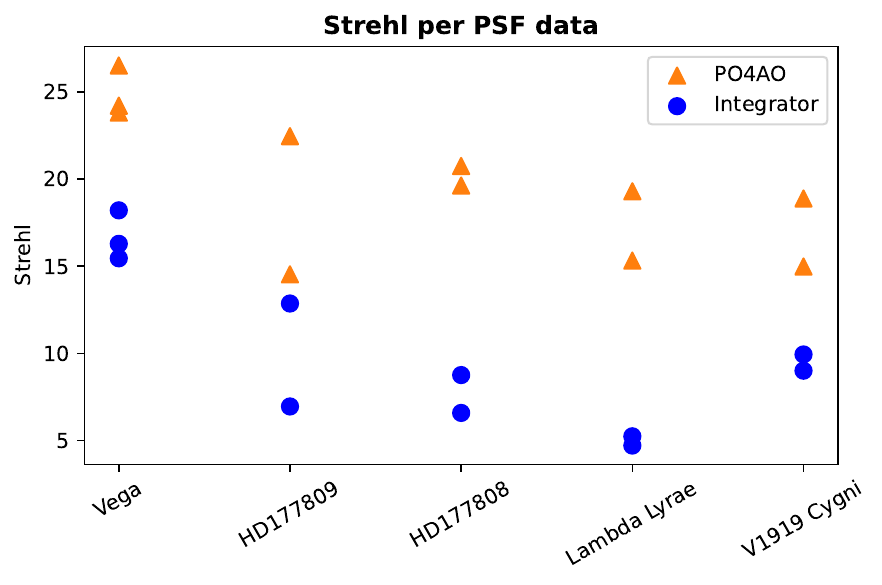}
    \caption{Strehl estimation of each PSF data set ($\sim$1min exposure) for selected targes. }
    \label{fig:all_second_night_psf}
\end{figure}

During the 2nd night, the vibration issue was fixed through tracking motor maintenance, and the PSF was more stable for both the integrator and PO4AO. We recorded data on multiple targets to assess PO4AO's robustness, performance against photon noise, and turnkey nature: Vega (0.09), Lambda Lyrae (4.93), HD177809 (5.72), HD177808 (6.13), Cygni (6.66). The Strehl estimates of all PSF data sets on selected targets are presented in Fig. \ref{fig:all_second_night_psf}. The Strehl ratio provided by PO4AO is always better than the best one provided by the integrator. However, the integrator was run at a fixed gain of 0.4, which is a good compromise between loop robustness and performance. Applying a higher or lower gain could have improved performance for bright and faint stars, respectively. The consistently higher performance across a wide range of guide-star magnitudes demonstrates PO4AO's robustness and ability to automatically adapt to prevailing conditions.

\begin{figure}[ht]
\centering
\subfloat[\label{fig:vega_var}]
        {\includegraphics[trim={0.cm 0cm 0cm 0cm}, clip, width=0.45\textwidth]{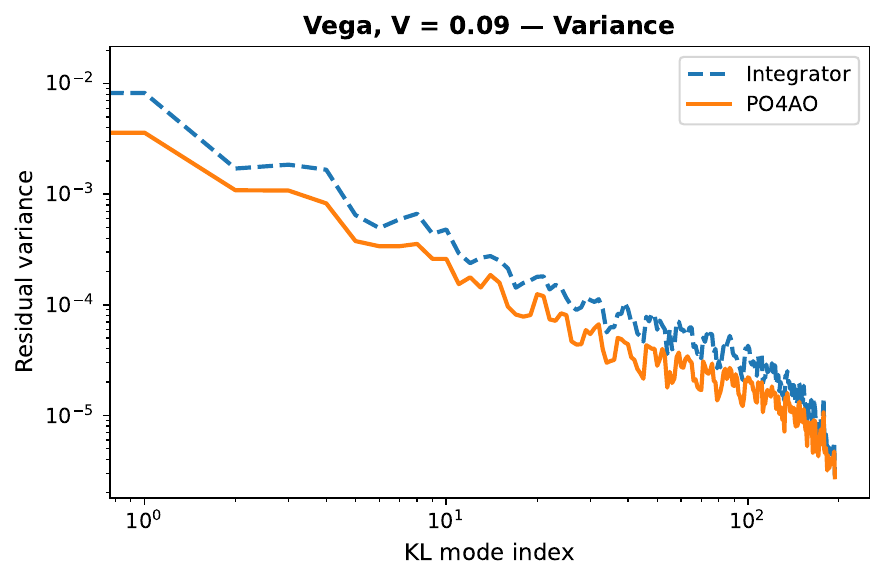}}

\subfloat[ \label{fig:HD_var}]
         {\includegraphics[trim={0.cm 0cm 0cm 0cm}, clip, width=0.45\textwidth]{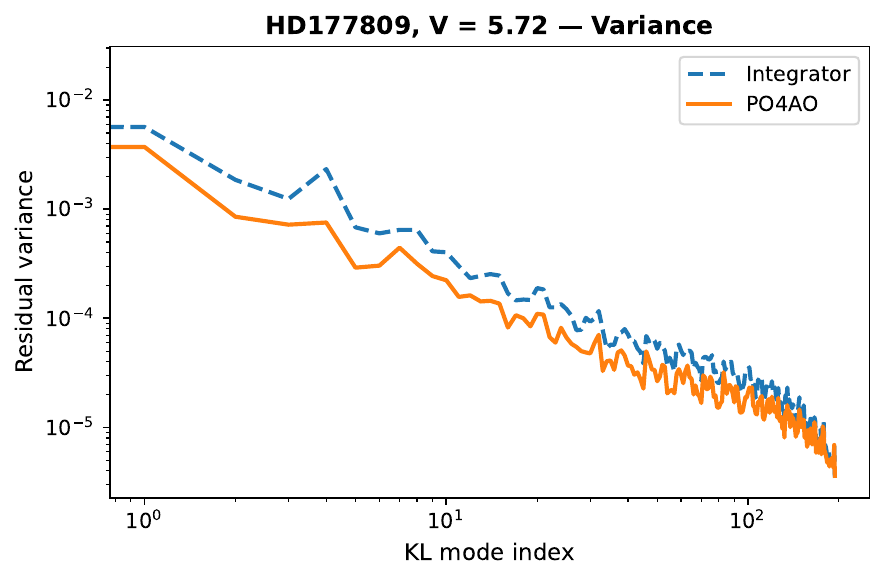}}

\subfloat[ \label{fig:cygni_var}]
         {\includegraphics[trim={0.cm 0cm 0cm 0cm}, clip, width=0.45\textwidth]{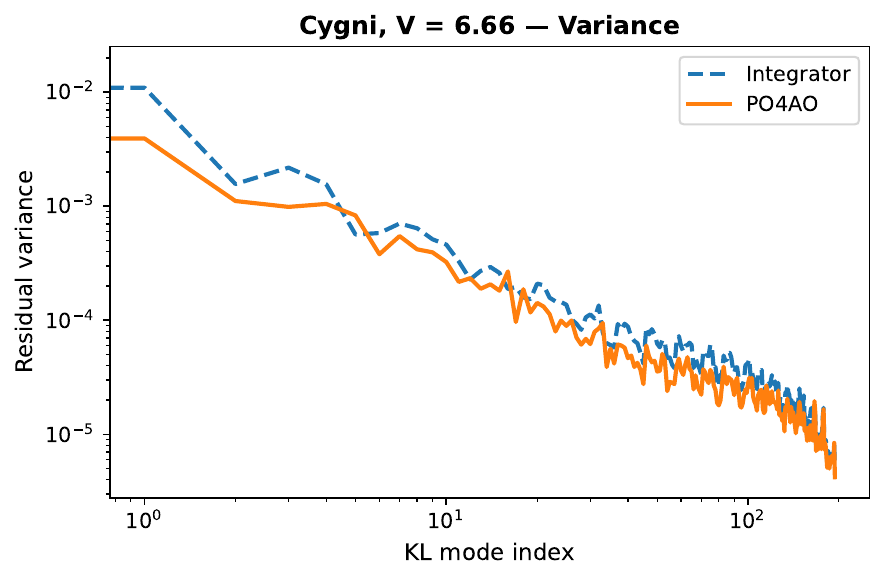}}

\caption[Training]
        {Second night telemetry analysis: Vega modal variance (a), Vega PSD of a "tilt" (b),  HD177809 modal variance (c), The orange lines correspond to PO4AO and the dashed blue lines to the integrator. The integrator was not properly tuned for each target and was therefore suboptimal.}
    \label{fig:2nd_telemetry}
\end{figure}

We selected the best PSF dataset from both methods for further analysis. We compute the long-exposure PSF (Figs. \ref{fig:second_night_psf}), peak-intensity comparisons (Table \ref{table:results}), and the modal variance from the telemetry data \ref{fig:2nd_telemetry}. PO4AO provides a higher Strehl ratio across all cases, and its PSFs are visibly sharper than the integrator PSFs. Additionally, the peak-intensity comparison aligns with the Strehl estimates, as shown in Table \ref{table:results}. We note that for the target "Vega," we corrected the NCPA errors; for the other targets, the NCPA is not corrected (hence the strong astigmatism).

Figures \ref{fig:2nd_telemetry} (a, c, e) show the residual modal variance. PO4OA again improves the performance across almost all KL modes for all targets. PO4AO yields a significant reduction in variance, particularly for lower-order modes, with gains up to a factor of 3.5. Mostly, the gain in modal variance is between 1.5 and 2 (see the easier-to-read plots in the Appendix). 

\subsection{Third night,  9/11/2025: comparing to DAO integrator}\label{sec:3rd}

\begin{figure}
\centering
\subfloat[\label{fig:3rd_psf}]
        {\includegraphics[trim={0.cm 0cm 0cm 0cm}, clip, width=0.48\textwidth]{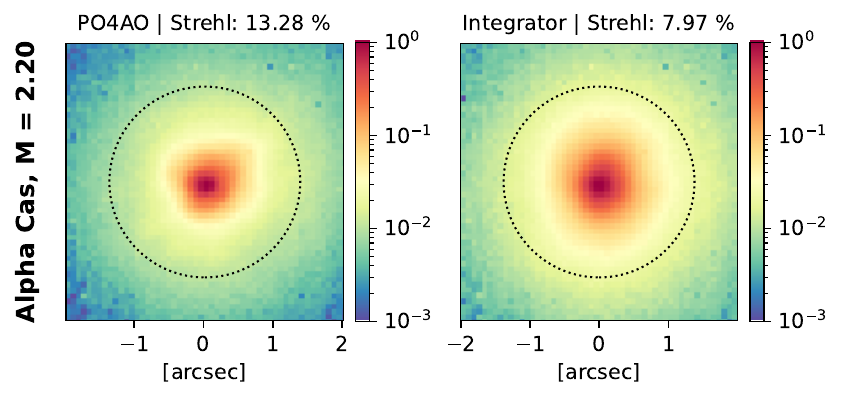}}

\subfloat[\label{fig:3rd_psf2}]
         {\includegraphics[trim={0.cm 0cm 0cm 0cm}, clip, width=0.48\textwidth]{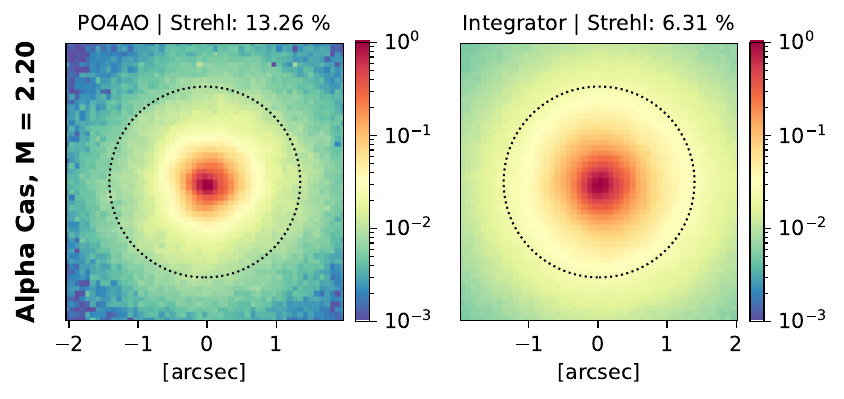}}
     
\caption[First night]
        {PSFs during the 3rd night, under Bad seeing. Left images: PO4AO (13.28\% and 13.26\% Strehl), Right image: DAO Integrator (7.97.46\% and 6.31\% Strehl)}
    \label{fig:third_night_psfs}
\end{figure}

\begin{figure}
\centering
\includegraphics[trim={0.cm 0cm 0cm 0cm}, clip, width=0.45\textwidth]{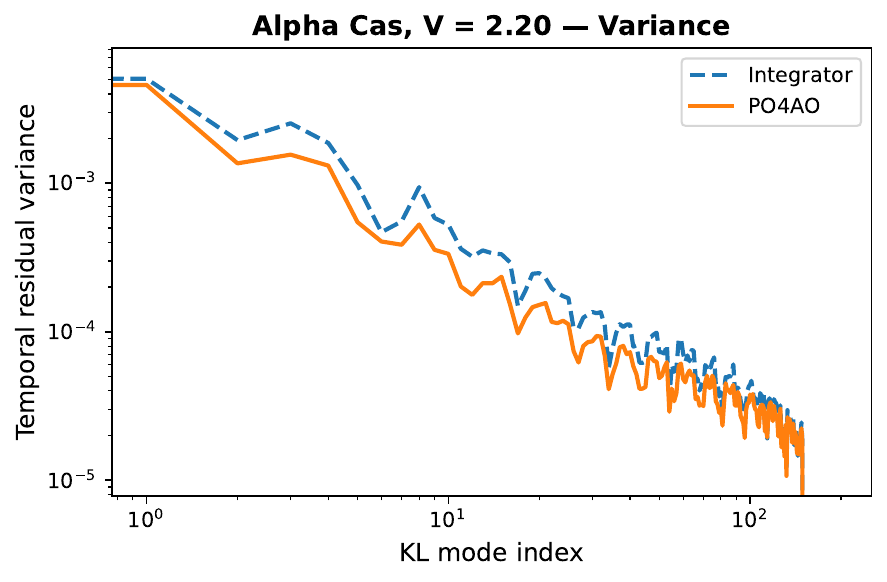}
\caption[First night]
        {Third night telemetry analysis.}
    \label{fig:3rd_telemetry}
\end{figure}

To validate the results from our first mission (the 1st and 2nd nights), we conducted additional tests several months later. This time, we compared PO4AO with the DAO internal integrator, after manually tuning the gain prior to recording. The seeing at night was around four arcsec. Again, we plot the PSF and analyze the telemetry. For both methods, we recorded two PSF data sets.

Fig. \ref{fig:third_night_psfs} shows the PSFs and Fig. \ref{fig:3rd_telemetry} the telemetry analysis of the better performing data sets. PO4AO delivers higher Strehl and peak intensity than the DAO integrator with optimized gain; see Table \ref{table:results}. Further, we compute the modal variance and temporal PSD of tilt. This time, the reduction in the modes' variance is less pronounced, from 1.1 to 2. The telemetry data do not show a significant improvement; however, we see a significant difference in PSFs (sharpness and Strehl ratio). This discrepancy we could not explain.
 
\begin{table}
    \caption{Strehl values and relative peak intensities (relative to PO4AO)}
    \label{table:results}
\begin{tabular}{ l l l l} 
 \hline\hline
 \multicolumn{4}{c}{First night} \\
 \hline
         Target & Method  & Strehl & relative peak intensity \\
 \hline
 Vega & PO4AO &  12.1\%  &   -   \\
 Vega & Python Int.  &  10.5\%  &  0.77     \\
 \hline
  \multicolumn{4}{c}{Second night} \\
 \hline
 Vega & PO4AO &  26.5\%  &   -   \\
 Vega & Python Int.  &  18.2\% &     0.74     \\
 HD177809 & PO4AO &  22.45\%  &  -    \\
 HD177809 & Python Int.  &  12.86\%  & 0.68         \\
 Cygni & PO4AO &  18.9\%  &    -  \\
 Cygni & Python Int.  &  9.9\%  &     0.60    \\
 \hline
  \multicolumn{4}{c}{Third night} \\
 \hline
 $\alpha$ Cas & PO4AO &  13.3\%  & -     \\
 $\alpha$ Cas & DAO Int.  &  8.0\%  & 0.62     \\
 \hline
\end{tabular}
\end{table}

\section{Linear analysis of behavior}\label{sec:analysis}

\subsection{Linear control matrix approach}

The purpose of this section is to highlight some features of how PO4AO uses measurements to close the control loop during the PAPYRUS on-sky experiment, thereby providing insight into the control law it implements. To this end, we analyze the PO4AO controller by approximating it as a linear controller, which allows for a tractable theoretical analysis. For the sake of simplicity, and in order to highlight the differences with the classical integrator controller, we consider a framework in which the controller only computes the commands sent to the deformable mirror using the measured aberrations at times $t$ and $t-1$ (during the PAPYRUS on-sky run, PO4AO operated using a history of 64 frames, $t$ and $t-1$ beeing the most informative ones). In both the PO4AO and integrator cases, the commands are incrementally added to the previous deformable mirror commands, with the inclusion of a leak factor $\alpha$. In a general form, the control law can be written as:

\begin{equation}
    \bm{v}_{t} = \alpha \cdot \bm{v}_{t-1} + \hat{C} \begin{bmatrix}
\bm{o}_{t} \\
\bm{o}_{t-1}
\end{bmatrix}
\label{eq:dm_cmds}
\end{equation}

with $\bm{v} \in \mathbb{R}^{n_{act}}$ the commands sent to the DM, $\bm o \in \mathbb{R}^{n_{act}}$ the measured residual measurements (expressed in DM commands), and $\hat{C}  \in \mathbb{R}^{n_{act}\times 2n_{act}}$ what we call the command matrix of the system. The goal of this analysis is to estimate the $\hat{C}$ matrix from the AO telemetry (which provides the quantities $\bm{D}$ and $\bm o$ at each time step). From Eq. \eqref{eq:dm_cmds}, we can write that the residual commands $\bm{a}$ sent to the DM at timestep $t$ are:

\begin{equation}
     \bm{a}^{t} = \hat{C} \begin{bmatrix}
\bm{o}_{t} \\
\bm{o}_{t-1}
\end{bmatrix} = \bm{v}_{t} - \alpha \cdot \bm{v}_{t-1}
\end{equation}

To retrieve the control matrix $\hat{C}$, one can build two history matrices $\mathbf{H}^{s} \in \mathbb{R}^{n_{act}\times l}$ and $\mathbf{H} \in \mathbb{R}^{2n_{act}\times l} $ from $l$ timesteps, such as:

\begin{equation}
\begin{aligned}
&\mathbf{H}^{s} =
\begin{bmatrix}
\bm{a}(1) & \cdots & \bm{a}(t) & \cdots & \bm{a}(l) 
\end{bmatrix}\\ \\
&\mathbf{H} =
\begin{bmatrix}
\bm {o}(1) & \cdots & \bm{o}(t) & \cdots & \bm{o}(l) \\
\bm{o}(0) & \cdots & \bm{o}(t-1) & \cdots & \bm{o}(l-1)
\end{bmatrix}
\end{aligned}
\end{equation}

Using these history matrices, we can estimate the control matrix:

\begin{equation}
     \hat{C} = \mathbf{H}^{s}\mathbf{H}^{+}
     \label{eq:history}
\end{equation}

where $\cdot^{+}$ denotes the pseudo-inverse.

\subsection{Application on PAPYRUS telemetry}

\begin{figure}
    \centering
    \includegraphics[width=0.98\linewidth]{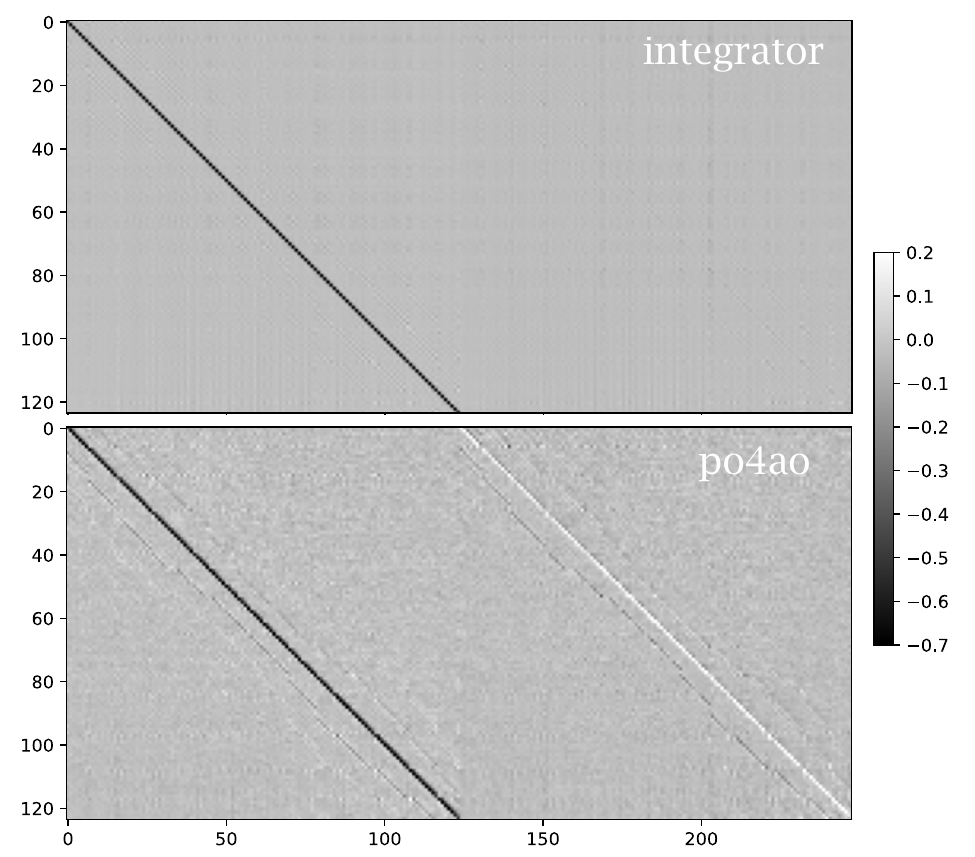}
    \caption{Control matrices $\hat{C}$ for the integrator and po4ao controllers. \textbf{Top:} Integrator. \textbf{Bottom:} PO4AO.}
    \label{fig:control_matrices}
\end{figure}

Using PAPYRUS telemetry acquired with the integrator controller and the PO4AO closed-loop configuration described in Sec. ~\ref{sec:po4ao}, we computed the control matrix $\hat{C}$ for each case. The computation used $l = 15,000$ timesteps and included only the actuators fully illuminated on the deformable mirror during on-sky operations, corresponding to $n_{\mathrm{act}} = 124$ actuators. The resulting control matrices are shown in Fig.~\ref{fig:control_matrices}.

As expected, the integrator command matrix exhibits the characteristic behavior of an integrator controller: the command applied to each actuator on the deformable mirror at time $t$ is obtained by multiplying the reconstructed actuator value at time $t$ by a negative loop gain. No dependence on the measurement at time $t-1$ is observed, in accordance with the integrator control law and the loop gain that was used during the closed loop:

\begin{equation}
\mathbf{a}_{t} = -0.7 \cdot \bm{o}_{t}.
\end{equation}

As a side note, the relatively high loop gain may seem surprising; however, it is intentionally chosen to partially compensate for the optical gain of the pyramid wavefront sensor.

\begin{figure}
    \centering
    \includegraphics[width=0.98\linewidth]{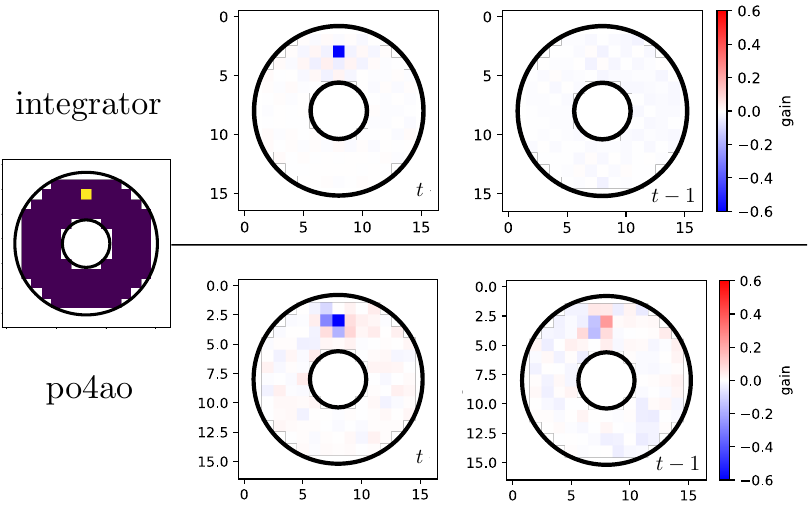}
    \caption{Maps of actuator contributions to the command sent to a given actuator. \textbf{Top:} Integrator for time steps $t$ and $t-1$. \textbf{Bottom:} PO4AO for time steps $t$ and $t-1$. These maps suggest the predictive nature of the PO4AO.}
    \label{fig:actuators}
\end{figure}

Still referring to fig.~\ref{fig:control_matrices}, we observe that the PO4AO control matrix exhibits a different structure. Within each $n_{act} \times n_{act}$ block, a larger number of off-diagonal terms is present, indicating that the command computed for a given actuator depends not only on the measurement at its own location, but also on measurements from other actuators. In addition, the matrix reveals dependence on measurements acquired not only at time $t$, but also at time $t-1$. We observe a positive feedback signature at the actuator location at time $t-1$ and a negative one at time step $t$. The lag forces the system into a pursue-and-correct cycle, naturally imprinting opposite signs on consecutive time-series coefficients. To better identify which actuators contribute to the control-law computation, Fig.~\ref{fig:actuators} shows, for a given actuator, the two columns of $\mathbf{C}$ corresponding to the contributions of actuator measurements at times $t$ and $t-1$. Once again, the integrator behaves as expected: the command sent to a given actuator is proportional to the measured value of that same actuator, scaled by the loop gain (with non-zero coefficients arising from the pseudo-inverse computation). In contrast, the PO4AO case shows that the command computed for a given actuator depends not only on its own measurement, but also on those of neighboring actuators, with a preferential orientation that likely corresponds to the wind direction. This behavior suggests the PO4AO controller's predictive nature.

\begin{figure}[h]
    \centering
    \includegraphics[width=1\linewidth]{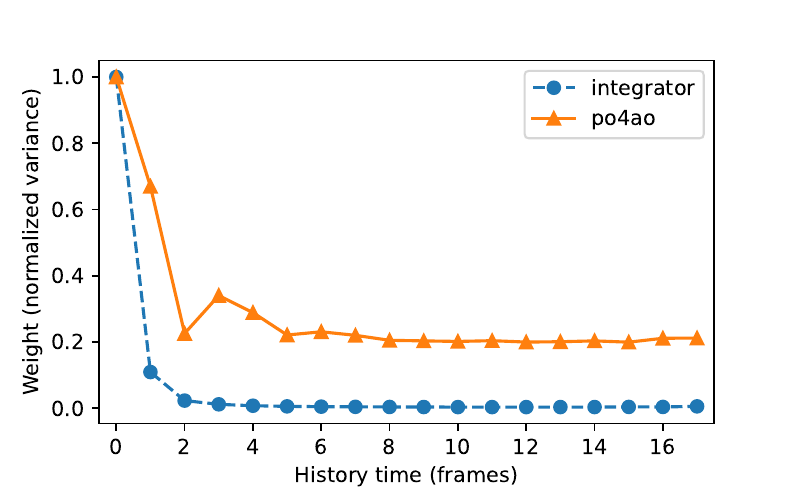}
    \caption{Analysis of the contribution of each history command. The weights are computed as the sum of the variance of all commands, normalized by the weight at timestep 0. }
    \label{fig:time_weight}
\end{figure}

One can also investigate how multiple past time steps contribute to the commands sent to the DM. To this end, we extend the linear analysis to include a longer temporal history, considering here up to 18 previous time steps (this number is limited by the increasing number of missing frames, which significantly reduces the available sample size). The command matrix therefore becomes $\hat{C} \in \mathbb{R}^{n_{\mathrm{act}} \times 18 n_{\mathrm{act}}}$. To quantify the contribution of each temporal lag, we define, for the $k^{\mathrm{th}}$ time step, the quantity $W(k)$ as the sum for all actuators of the variance of their contribution function:

\begin{equation}
    W(k) = \sum_{i=0}^{N_{act}}\sum_{j=0}^{N_{act}}\hat{C}(i,k\times N_{act} + j)^2
\end{equation}
This quantity is normalized by its value at time step $k=0$ and displayed in Figure~\ref{fig:time_weight} for each timestep. As previously observed, the integrator controller is dominated almost entirely by the contribution from the current time step ($k=0$, other timestep contributions are due to the imperfect inversion of the history matrix), whereas the PO4AO controller exhibits significant contributions from several previous time steps.

We note that the weights associated with the PO4AO controller reach a non-zero floor, attributable to noise in the estimated command matrix. This residual floor likely reflects the fact that the controller cannot be perfectly represented by a purely linear model, resulting in an imperfect reconstruction by the linear approximation. For the PO4AO controller, the dominant contributions appear to arise primarily from time steps up to approximately $k=8$. Although this temporal horizon remains relatively short, it may suggest that the temporal error is largely driven by high-speed frozen-flow turbulence, which can be effectively predicted using only a few previous time steps. Further, this analysis was performed on a single data set from the third night and only reflects the required history length for that observation. We tuned the history length on the first night all the way to 64 frames. \cite{nousiainen2024laboratory} showed that vibration cancellation benefits from longer history lengths, and this extended history length of 64 could be due to the strong vibration during the first night.

\begin{figure}
    \centering
    \includegraphics[width=0.9\linewidth]{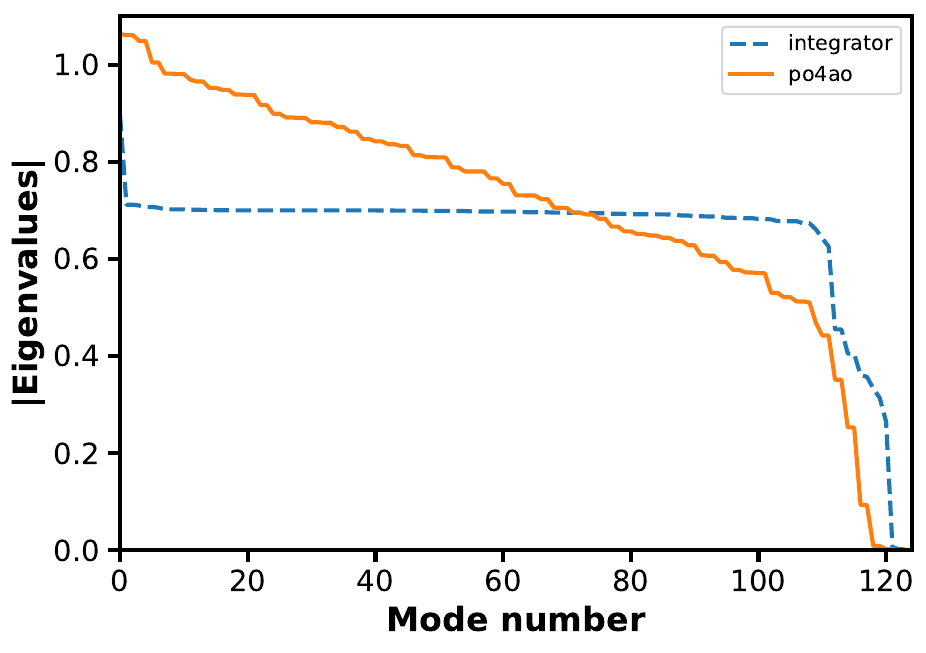}
    \caption{Eigenvalues for the first $n_{act} \times n_{act}$ block of the control matrices of the integrator and PO4AO controllers.}
    \label{fig:eigenvalues}
\end{figure}

Another way to highlight the more complex control law implemented by PO4AO is to analyze the properties of certain modes of the control matrix. To simplify this analysis, we ignore the dependence on $t-1$ and consider only the first $n_{act} \times n_{act}$ block of the control matrices (while this is not a fully rigorous approach, it provides a convenient representation to visualize and compare the behavior of different modes). Fig. ~\ref{fig:eigenvalues} shows the eigenvalues of these blocks. For the integrator, nearly all modes exhibit a uniform value corresponding to the loop gain, with deviations for higher-order modes arising from the imperfect inversion of the history matrix $\mathbf{H}$ in Eq.~\eqref{eq:history}. In contrast, the PO4AO eigenmodes exhibit a range of gains, as expected for a more sophisticated controller that optimizes the gains based on the controlled modes.

\section{Conclusions and future work}\label{sec:conclusion}
In this paper, we presented the first on-sky experimental validation of the reinforcement learning controller PO4AO on the Papyrus adaptive optics system, with particular emphasis on its practical real-time implementation and operational behavior. When properly implemented and tuned, PO4AO can serve as a robust, turnkey controller for single-conjugate AO systems running on-sky. Its ability to operate without expert intervention with a fixed set of hyperparameters across diverse observing conditions and targets makes it a promising candidate as the controller for operations in which AO performance is crucial, such as high-contrast imaging \citep{guyon2018extreme, guyon2005limits}. 

Our experiments show that PO4AO delivers consistent performance gains over the Python-based single-gain-integrator across all tested scenarios, including varying seeing conditions, target brightness, vibration patterns, and wind velocity. We note here that the Python-based integrator gain was not always at the optimal value, and no modal gain values were implemented. Hence, the performance gain of PO4AO can be attributed to several factors. To this end, we analyzed the telemetry, which suggested vibration rejection and robustness to photon noise (i.e., robustness to fainter guide stars). We also conducted a brief linear analysis of PO4AO behavior. The behavior follows patterns one would expect from an advanced AO control law: it treats spatial information non-uniformly, utilizes past telemetry, suggesting adaptation for wind-driven prediction, and, also, differentiates eigenmodes with gain values that indicate adaptation to the reconstruction matrix's noise propagation. Moreover, PO4AO also seemed to outperform a well-tuned low-latency DAO integrator; however, due to some corrupted data and limited testing time, confirming this result would require additional experiments over multiple nights. Also, due to challenging seeing conditions and significant frame-skipping in the telemetry data, a more thorough analysis of the error terms is not feasible.

The current implementation adds an additional $720 \pm 280$ \textmu s to the pipeline latency, making it suitable for running Papyrus around 500Hz. During the test, Papyrus had only one GPU card; training and control were performed simultaneously on the same card, resulting in memory-bandwidth overheads. This led to fluctuations (depending on the training circle position) in control jitter and consequently to frame skipping (around 5\% and sometimes several at once). PO4AO remained stable despite this non-optimal implementation and even seemed to outperform the low-latency DAO integrator.

Establishing PO4AO as the standard control model for Papyrus would require further development. First, the code needs to be refactored and integrated with a graphical user interface to enhance accessibility for non-expert users. Second, an additional GPU card should be installed to handle the training thread and ensure stable hard real-time performance. Third, PO4AO would greatly benefit from re-implementing critical code sections in lower-level programming languages such as C, enabling direct memory access and optimized execution. Furthermore, the neural network could be deployed with NVIDIA TensorRT, a high-performance inference optimizer that delivers low latency and high throughput for deep learning applications. This would significantly reduce latency and jitter, potentially further enhancing PO4AO performance-- substantial, non-uniform jitter and latency make predictive control difficult. 

Future work will focus on deploying PO4AO on high-contrast imaging instruments at very large-scale telescopes, such as MagAO-X (Magellan Adaptive Optics eXtreme system, \citealp{males2018magao}) and SCExAO (Subaru Coronagraphic Extreme Adaptive Optics, \citealp{jovanovic2015subaru}), and the SAXO+ (upgrade to SAXO, the AO system of the VLT-SPHERE; \citealp{boccaletti2020sphere+}) project. Another promising direction is exploring the use of PO4AO for highly nonlinear wavefront sensing, including Zernike and non-modulated pyramid WFS. 

\begin{acknowledgements}
This work benefited from the support the French National Research Agency
(ANR) with the Programme Investissement Avenir F-CELT (ANR-21-ESRE-
0008), the ANR-DGA-AID ASTRID program (ANR-25-ASTR-0015), the PEPR ORIGINS, the
Action Spécifique Haute Résolution Angulaire (ASHRA) of CNRS/INSU320
co-funded by CNES, the french government under the France 2030 investment
plan (cassiopée project) and the Initiative d’Excellence d’Aix-Marseille
Université A*MIDEX, program number AMX-22-RE-AB-151
\end{acknowledgements}

\bibliographystyle{aa}
\bibliography{aanda}
\begin{appendix} 
\section{Additional telemetry analysis}
\begin{figure}[h]
\centering
\subfloat[\label{fig:vega_mse1}]
        {\includegraphics[trim={0.cm 0cm 0cm 0cm}, clip, width=0.45\textwidth]{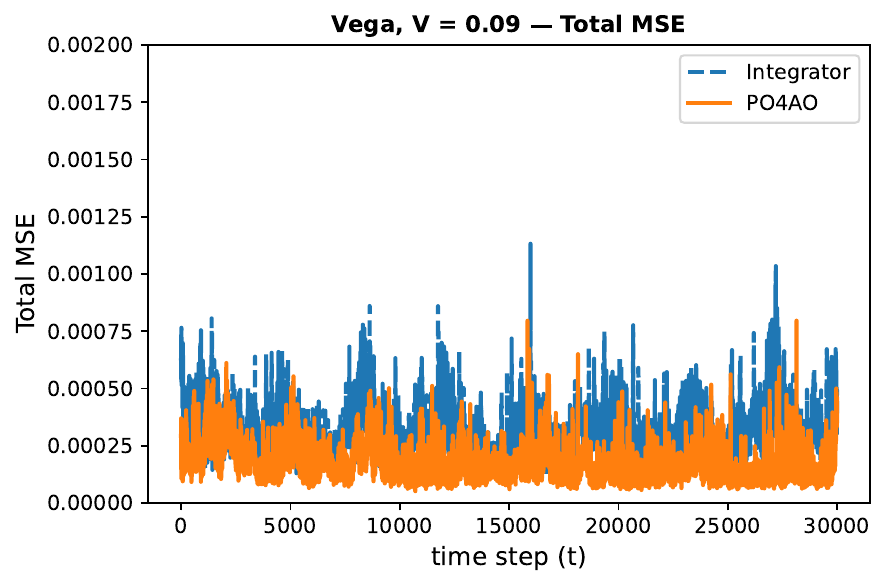}}
   
\subfloat[\label{fig:vega_var_gain}]
         {\includegraphics[trim={0.cm 0cm 0cm 0cm}, clip, width=0.45\textwidth]{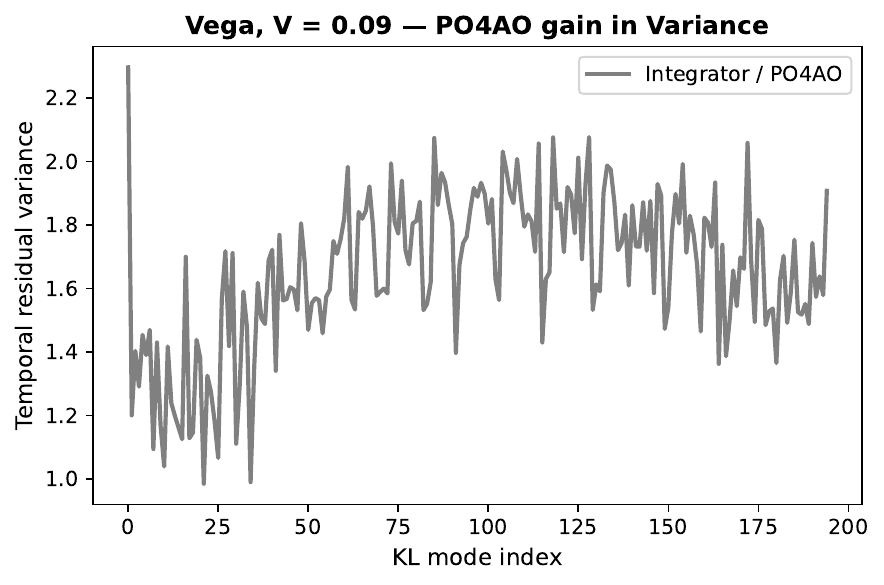}}  
      
\caption[add]
        {Additional first night telemetry analysis. (a) mean-squared wavefront error at each time step: blue line is for the integrator and orange line for PO4AO. (b) Comparison between residual modal variance, i.e., integrator variance divided by the PO4AO variance for each KL mode.  }
    \label{fig:1st_telemetry_add}
\end{figure}

For interested readers, we have added several additional telemetry plots. For each dataset presented in the paper, we plot the wavefront mean-squared error (MSE) at each time step and compare the modal variance of PO4AO to illustrate the relative gain discussed in the results section. Fig.~\ref{fig:vega_mse1} shows the results of the first night. The PO4AO delivers a lower overall MSE than the integrator. Fig.~\ref{fig:vega_var_gain} shows the earlier discussed gain in variance.

\begin{figure}[h]
\centering
\subfloat[\label{fig:vib_appendix}]
        {\includegraphics[trim={0.cm 0cm 0cm 0cm}, clip, width=0.45\textwidth]{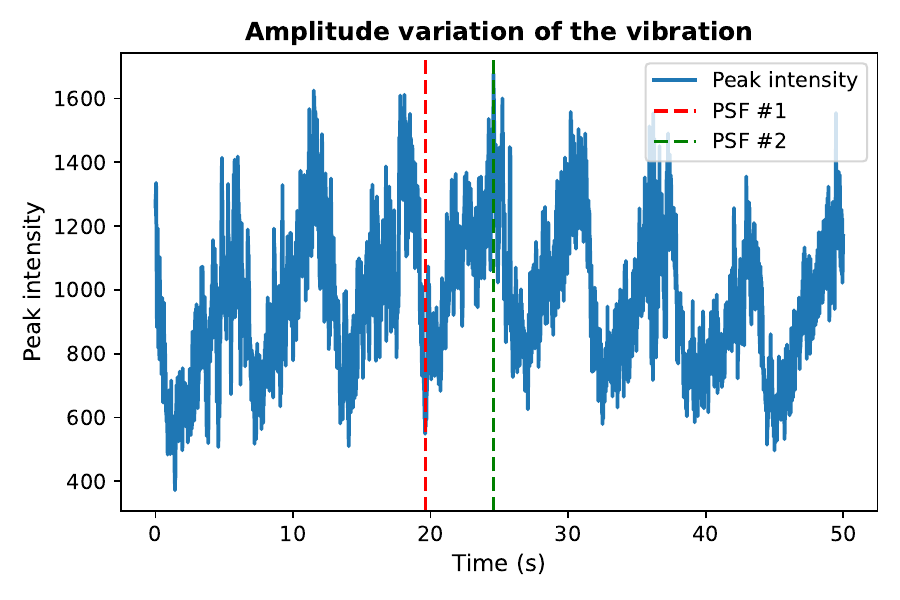}}
   
\subfloat[\label{fig:vib_appendix_psf}]
         {\includegraphics[trim={0.cm 0cm 0cm 0cm}, clip, width=0.45\textwidth]{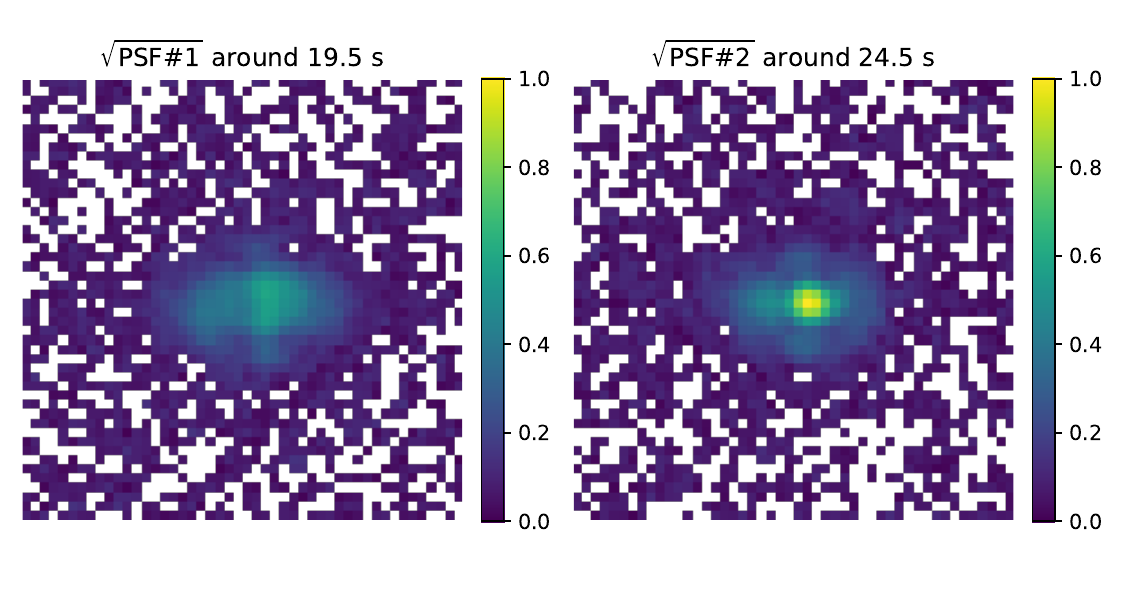}}  
      
\caption[add]
        {Analysis of the vibration during the first night. (a) The PSF peak intensity at each recorded integrator-controlled science camera image. The blue line shows the peak intensity, and the dashed vertical lines indicate the PSF locations. (b) a low- (PSF \#1, frame \#1950) and high-intensity (PSF \#2, frame \#2547) PSF}
    \label{fig:vibration_add}
\end{figure}

Moreover, to illustrate the non-uniform nature of the vibration observed on the first night, we compute the peak intensity of the science camera over every frame in Fig.~\ref{fig:vibration_add}. We see that the vibration strength is non-uniform and fluctuates approximately every 6 seconds (see Fig.~\ref{fig:vib_appendix}). When the vibration is strong, the PSF elongates in the direction of the vibration (hence, low peak intensity), whereas when the vibration is mild, the PSF is more focused (high peak intensity); see Fig.~\ref{fig:vib_appendix_psf}. We observe that when peak intensity is low, the PSF is elongated in the vertical direction, suggesting strong vibration (vibration frequency $>=$ camera framerate)

\begin{figure*}
\centering
\subfloat[\label{fig:vega_mse}]
        {\includegraphics[trim={0.cm 0cm 0cm 0cm}, clip, width=0.45\textwidth]{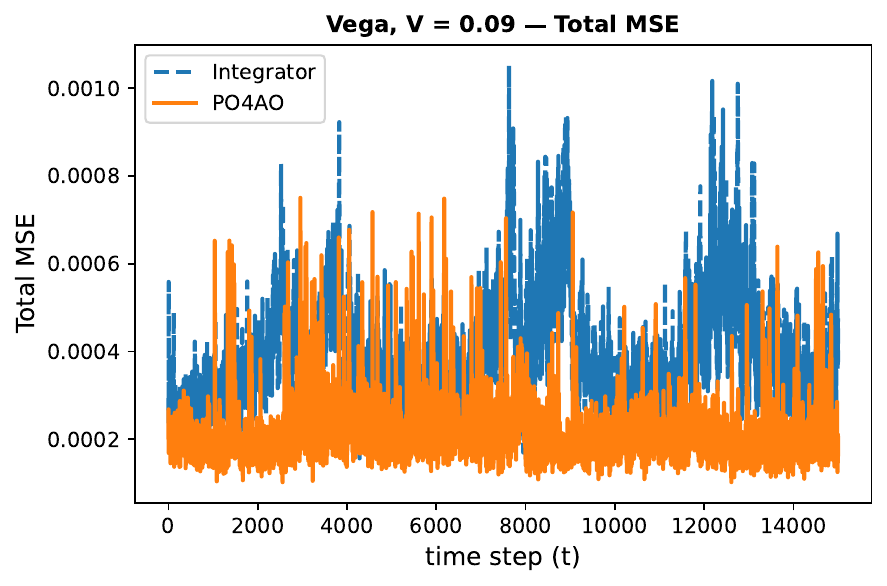}}
    \hfill
\subfloat[\label{fig:vega_gain}]
         {\includegraphics[trim={0.cm 0cm 0cm 0cm}, clip, width=0.45\textwidth]{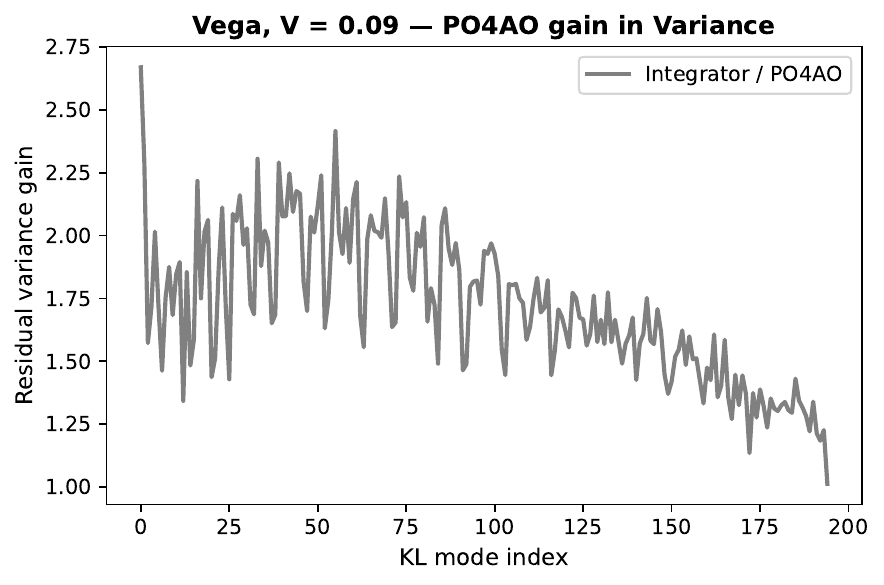}}

\subfloat[ \label{fig:HD_mse}]
         {\includegraphics[trim={0.cm 0cm 0cm 0cm}, clip, width=0.45\textwidth]{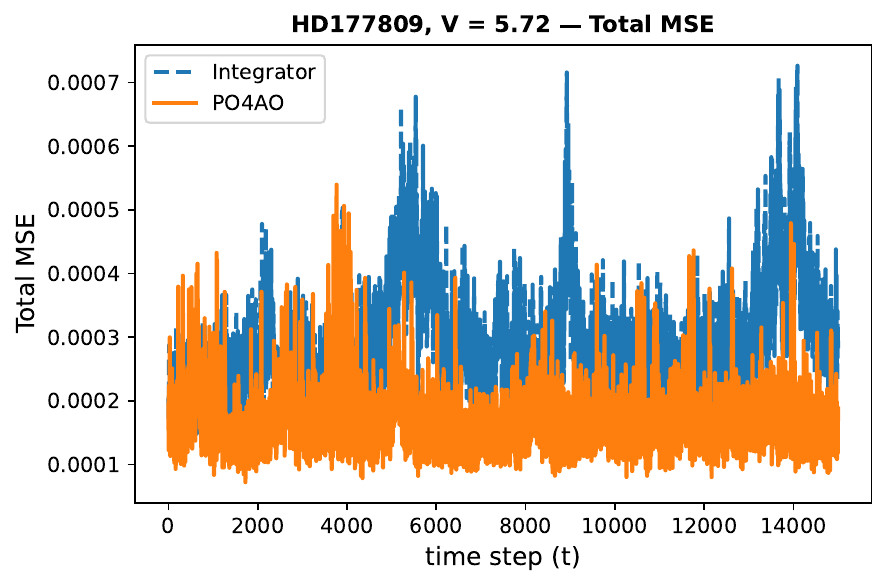}}
    \hfill
\subfloat[ \label{fig:HD_gain}]
        {\includegraphics[trim={0.cm 0cm 0cm 0cm}, clip, width=0.45\textwidth]{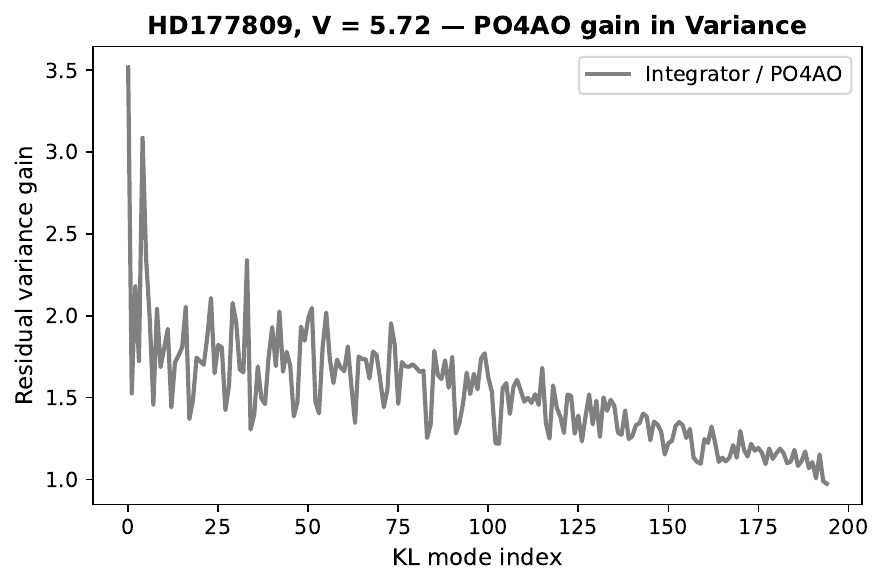}}

\subfloat[ \label{fig:cygni_mse}]
         {\includegraphics[trim={0.cm 0cm 0cm 0cm}, clip, width=0.45\textwidth]{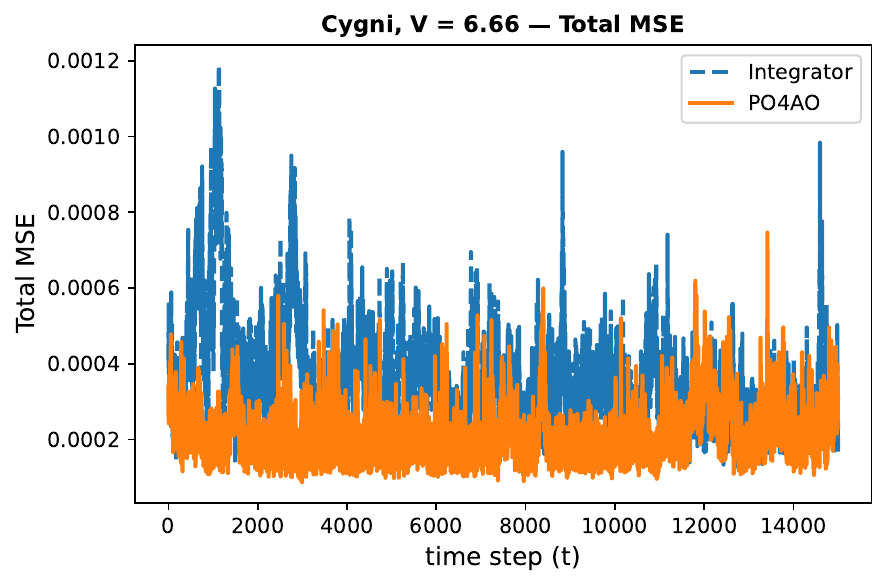}}
    \hfill
\subfloat[ \label{fig:cygni_gain}]
        {\includegraphics[trim={0.cm 0cm 0cm 0cm}, clip, width=0.45\textwidth]{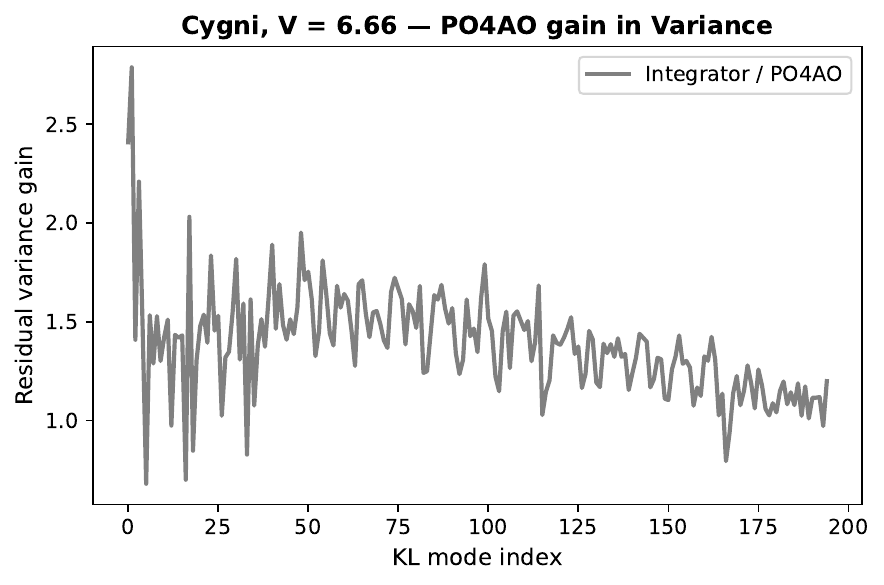}}        
      
\caption[add]
        {Additional second night telemetry analysis. (a, c, e) mean-squared wavefront error at each time step for each target: blue lines are for the integrator and orange lines for PO4AO. (b, d, f) Comparison between residual modal variance, i.e., integrator variance divided by the PO4AO variance for each KL mode for each target.  }
    \label{fig:2nd_telemetry_add} 
\end{figure*}

\begin{figure*}
\centering
\subfloat[\label{fig:alphacas_mse}]
        {\includegraphics[trim={0.cm 0cm 0cm 0cm}, clip, width=0.45\textwidth]{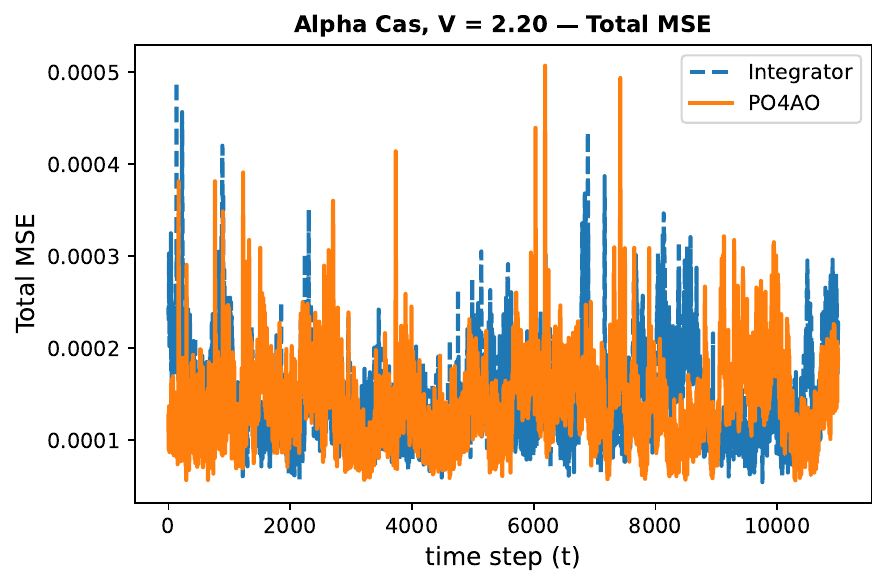}}
    \hfill
\subfloat[\label{fig:alphacas_gain}]
         {\includegraphics[trim={0.cm 0cm 0cm 0cm}, clip, width=0.45\textwidth]{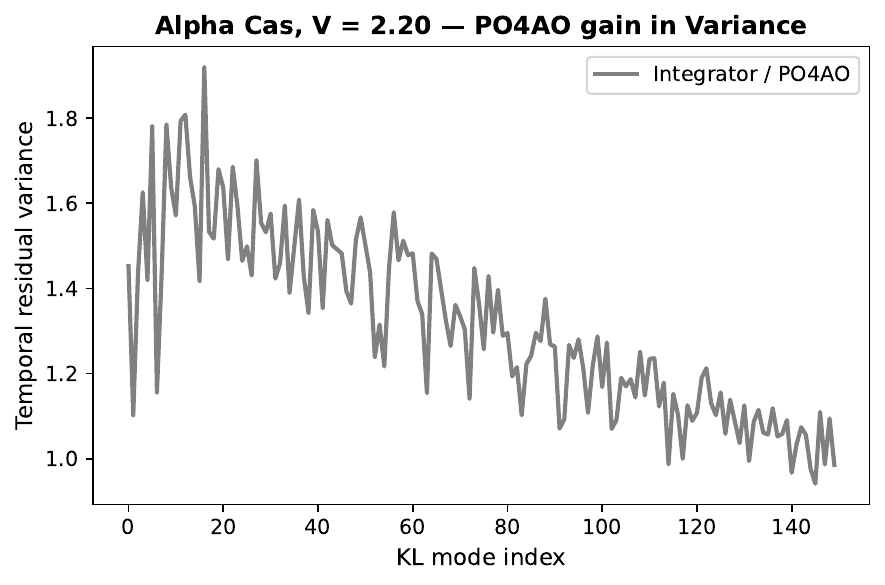}}  
      
\caption[add]
        {Additional third night telemetry analysis. (a) mean-squared wavefront error at each time step: blue line is for the integrator and orange line for PO4AO. (b) Comparison between residual modal variance, that is, integrator variance divided by the PO4AO variance for each KL mode.  }
    \label{fig:3rd_telemetry_add}
\end{figure*}

Finally, we also plot the additional telemetry analysis for the second and third nights. On the second night (Fig. \ref{fig:2nd_telemetry_add}), PO4AO again delivers clearly a lower wavefront MSE across all cases. On the third night (Fig. \ref{fig:3rd_telemetry_add}), the difference between the methods is not noticeable in the MSE. This may be due to challenging seeing conditions, which led to frequent DM saturation for both methods. 

\end{appendix}

\end{document}